\documentclass[11pt]{article}

\usepackage[margin=1in]{geometry}

\usepackage{graphicx}
\usepackage{amsmath, amssymb}
\usepackage{siunitx}
\usepackage{cite}
\usepackage{bm}
\usepackage{float}
\usepackage{caption}
\usepackage{subcaption}
\usepackage{hyperref}
\usepackage[left]{lineno}
\usepackage{xcolor}
\hypersetup{
    colorlinks=true,
    linkcolor=blue,
    citecolor=blue,
    urlcolor=blue,
}

\title{Multibit Ferroelectric Memcapacitor for Non-volatile Analogue Memory and Reconfigurable Filtering}
\author{Deepika Yadav$^{1,*}$, Spyros Stathopoulos$^1$, Patrick Foster$^1$, Andreas Tsiamis$^1$,\\ Mohamed Awadein$^1$, Hannah Levene$^1$, Themis Prodromakis$^{1,*}$\\
\\
\small $^1$Centre for Electronics Frontiers, School of Engineering, University of Edinburgh, UK\\
\small $^*$Correspondence: \texttt{deepika.yadav@ed.ac.uk}
}
\date{}

\begin{document}

\maketitle

\section*{Abstract}
Tuneable capacitors are vital for adaptive and reconfigurable electronics, yet existing approaches require continuous bias or mechanical actuation. Here we demonstrate a voltage-programmable ferroelectric memcapacitor based on HfZrO that achieves more than eight stable, reprogrammable capacitance states ($\sim$3-bit encoding) within a non-volatile window of $\sim$24~pF. The device switches at low voltages ($\pm$3~V), with each state exhibiting long retention ($>$10$^5$~s) and high endurance ($>$10$^6$ cycles), ensuring reliable multi-level operation. At the nanoscale, multistate charge retention was directly visualised using atomic force microscopy, confirming the robustness of individual states beyond macroscopic measurements. As a proof of concept, the capacitor was integrated into a high-pass filter, where the programmed capacitive states shift the cutoff frequency over $\sim$5~kHz, establishing circuit-level viability. This work demonstrates the feasibility of CMOS-compatible, non-volatile, analogue memory based on ferroelectric HfZrO, paving the way for adaptive RF filters, reconfigurable analogue front-ends, and neuromorphic electronics.

\section*{Introduction}
Capacitors are essential in modern electronics, supporting functions from sensing and actuation to analogue filtering and radio-frequency (RF) signal processing \cite{he_tunable_2020, Zhao2016}. Many of these applications demand tuneable capacitance. Conventional solutions include microelectromechanical systems (MEMS) and ferroelectric thin-film varactors (BaSrTiO$_3$, PZT). MEMS offer high-quality analogue tuning but are bulky due to mechanical actuation, while ferroelectric varactors suffer from poor CMOS compatibility \cite{Aziz2019ReconfigurableApplications,  mikolajick_next_2021, zhou_ferroelectric_2025}. These limitations have intensified interest in solid-state alternatives that combine compactness, low power, and reconfigurability. Memcapacitors, introduced as a theoretical extension of the memristor in Chua’s framework, address this need by offering history-dependent capacitance states that are programmable without static bias \cite{chua_memristor-missing_1971, DiVentra2009}. Several reports exist that display binary or multibit memcapacitive behaviour employing metal oxides or ferroelectrics in different frameworks, such as memristors, gate dielectrics, or metal-oxide-semiconductor stacks \cite{Park2018, yadav_impedance_2025, demasius_energy-efficient_2021, tian_ultralowpower_2023, zhang_flexible_2025}.  Among these, ferroelectric hafnium zirconium oxide (HZO) has emerged as a powerful candidate due to its robust orthorhombic phase \cite{muller_ferroelectric_2011,park_ferroelectricity_2015, pesic_physical_2016}. Stable binary polarisation, scalability to nanometer thickness, endurance, and CMOS compatibility have already enabled HZO as non-volatile memories such as FeRAM, FTJs, and FeFETs \cite{chen_working_2025,haglund-peterson_nonvolatile_2024,halter_multi-timescale_2023,luo_highly_2020,zhou_metal-insulator-semiconductor_2020,si_ferroelectric_2019}. Beyond polarisation-based memory, the characteristic butterfly-shaped capacitance–voltage response of HZO points to its potential for capacitive switching \cite{koduru_small-signal_2025}. Recent studies on ferroelectric memcapacitors have therefore focused on three central aspects: achieving a sizable capacitive memory window (CMW) at zero bias to enable non-destructive readout (NDRO), realising stable non-volatile capacitive states, and maximising the number of accessible states at the lowest possible operating voltages. Numerous approaches ranging from symmetric to asymmetric electrode designs, as well as doping strategies using Aluminium, Lanthanum, and other elements, have highlighted the promise of HZO-based memcapacitors. However, these efforts often address only individual aspects, leaving a comprehensive demonstration to be established \cite{habibi_enhancing_2025,tran2017memcapacitivedeviceslogiccrossbar, zhou_metal-insulator-semiconductor_2020,zhang_flexible_2025, zhou_ferroelectric_2025}. 

Building on this motivation, we present wake-up–free switching and robust binary states with NDRO in the optimal HZO stack configuration. The optimised stack exhibited voltage-controlled reprogrammable states with good retention and endurance. The observed capacitive window is larger than previously reported stacks \cite{habibi_enhancing_2025,mukherjee_capacitive_2023,wang_hafnium_2025}. To confirm that both binary and multibit capacitive states can be locally programmed and retained, electrostatic force microscopy was performed, providing nanoscale evidence for memcapacitive operation. Finally, to demonstrate circuit-level functionality, the ferroelectric capacitor was integrated into a tuneable RC high-pass filter showing frequency programmability. This material-to-device proof-of-concept demonstrates that ferroelectric memcapacitors can directly translate polarisation-defined states into functional, reconfigurable circuit responses. Together, these results position HZO-based memcapacitors as promising building blocks for analogue memory, RF tuning, and adaptive on-chip hardware.

\section*{Results}
\begin{figure*}[t]
\centering
\includegraphics[width=\textwidth]{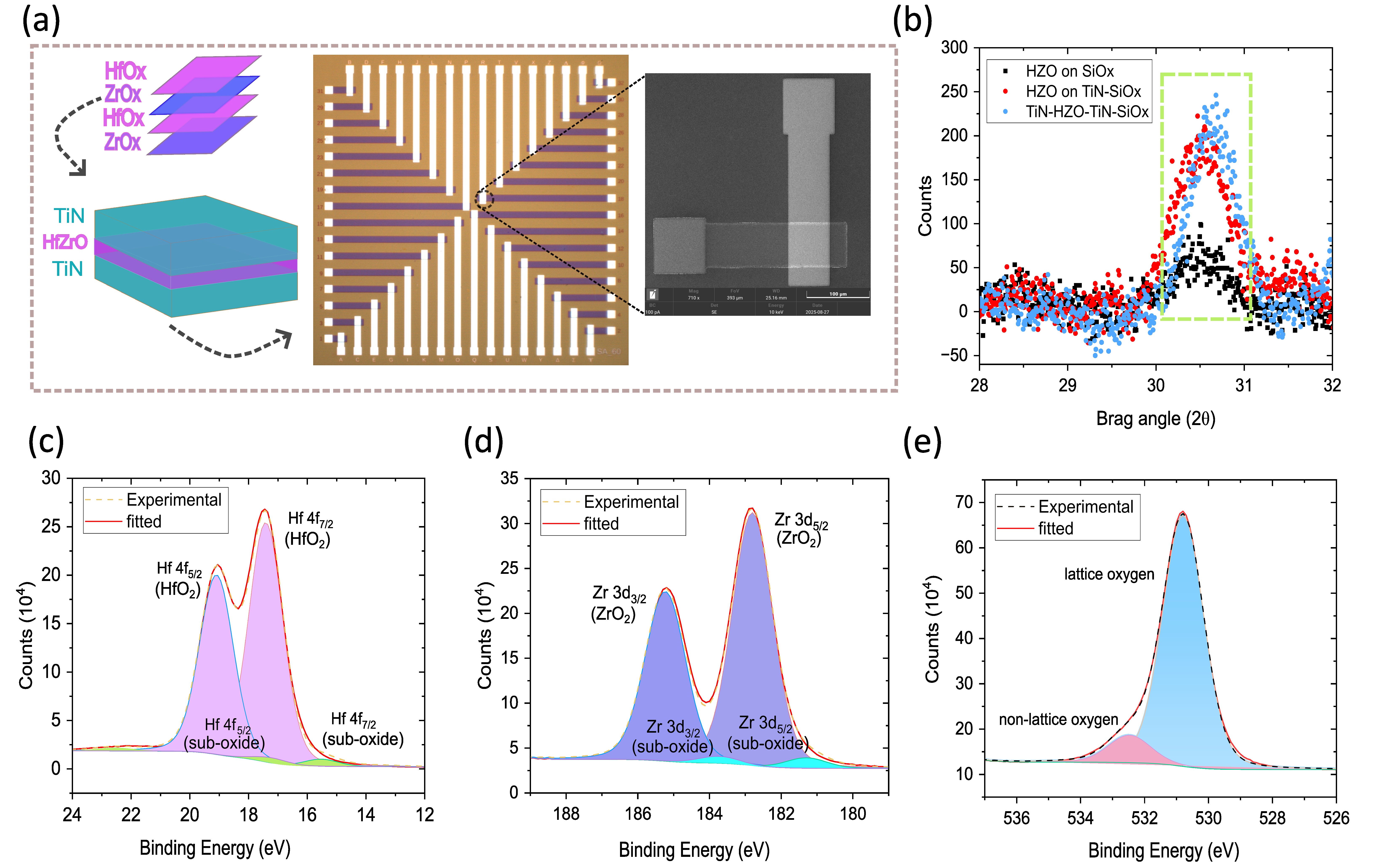}
\caption{\textbf{Device structure and film characterisation.} 
(a) Wafer schematic showing 32 standalone TiN/HfZrO/TiN capacitors, where the HfZrO layer is deposited by atomic layer deposition using alternating Hf and Zr precursors. Also, a SEM top view of an individual device highlights the electrode geometry and film uniformity. (b) Glancing angle XRD patterns for HZO on SiO$_x$, HZO on bottom TiN, and TiN/HZO/TiN stacks reveal a distinct orthorhombic phase peak near 30.5$^\circ$ that strengthens with TiN electrodes, indicating improved crystallinity and phase stabilisation. (c-e) XPS spectra of annealed HZO films show the expected Hf~4f and Zr~3d doublets and O~1s features, confirming correct cation valence and limited defect contributions. Together, these measurements verify the structural and chemical quality of the ferroelectric HZO layer.
}
\label{fig1}
\end{figure*}
We investigated standalone ferroelectric memcapacitors based on a TiN/HZO/TiN stack. The device layout, together with the HfO$\text{x}$/ZrO$\text{x}$ stacking sequence, is shown in Fig.~1(a), along with a scanning electron microscopy (SEM) image highlighting the nanoscale structure of an individual capacitor. To establish the material and device characteristics, we first present structural and chemical analyses of the HZO films using X-ray diffraction (XRD), X-ray photoelectron spectroscopy (XPS), and atomic force microscopy (AFM) [Figs. 1 and 2]. Electrical measurements then establish the polarisation dynamics and the existence of a capacitance memory window (CMW), followed by material optimisation to enhance device performance [Fig.~3]. These improvements enable demonstration of multibit operation [Fig.~4] and, subsequently, reconfigurable circuit functionality in a high-pass filter [Fig.~5].

\subsection*{Thin Film Properties}
Glancing-incidence XRD of the HZO films (Fig.~1b) shows a dominant peak at 2$\theta \approx$ 30.5°, characteristic of the ferroelectric orthorhombic phase \cite{kim_switching_2022}. Additional peaks (Fig.~1S(a-b)) confirm a polycrystalline structure with coexisting orthorhombic, tetragonal, and monoclinic phases, indicating that rapid thermal annealing (RTA)
stabilises ferroelectricity within a mixed-phase matrix. To assess the electrode effect, films with no TiN, only bottom TiN, and both top and bottom TiN electrodes were compared. The orthorhombic peak intensity increases progressively with electrode incorporation and shifts slightly to higher angles, indicating the role of TiN in suppressing the non-ferroelectric monoclinic phase. These results establish that TiN electrodes are critical for enhancing orthorhombic phase stability, which is consistent with earlier studies \cite{lee_influence_2021}. To investigate the stoichiometry of the HZO film and cation oxidation, XPS analysis was performed, and the results are presented in Figure~1(c–e). The Hf 4f spectrum (Fig.~1c) shows a sharp spin–orbit doublet, with Hf 4f$_\text{7/2}$ and Hf 4f$_\text{5/2}$ peaks at $\sim$17.2 eV and $\sim$18.8 eV, separated by $\sim$1.6 eV. Similarly, the Zr 3d spectrum (Fig.~1d) exhibits Zr 3d$_\text{5/2}$ and Zr 3d$_\text{3/2}$ peaks at $\sim$182.6 eV and $\sim$185.2 eV, with a splitting of $\sim$2.4 eV that is consistent with previously reported for HfO$_\text{x}$ and ZrO$_\text{x}$ films \cite{xiao_comparison_2020}. Quantitative fitting yields a near 1:1 Hf:Zr atomic ratio, as shown in Fig.~1S(c), a composition widely recognised as optimal for maximising ferroelectric polarisation in HZO. The Oxygen 1s spectrum (Fig.~1e) shows a dominant lattice-oxygen peak at $\sim$530.6~eV and a smaller feature at $\sim$532.3 eV, pointing to only modest oxygen defect density. The clean Hf$^{4+}$/Zr$^{4+}$ core levels, together with the low defect fraction, are consistent with the robust electrical behaviour. Importantly, the peak positions of Hf, Zr, and O remain essentially unchanged after rapid thermal annealing (RTA), showing that annealing preserves cation valence states while facilitating structural evolution toward the orthorhombic ferroelectric phase.

\begin{figure*}[t]
\centerline{\includegraphics[width=\textwidth]{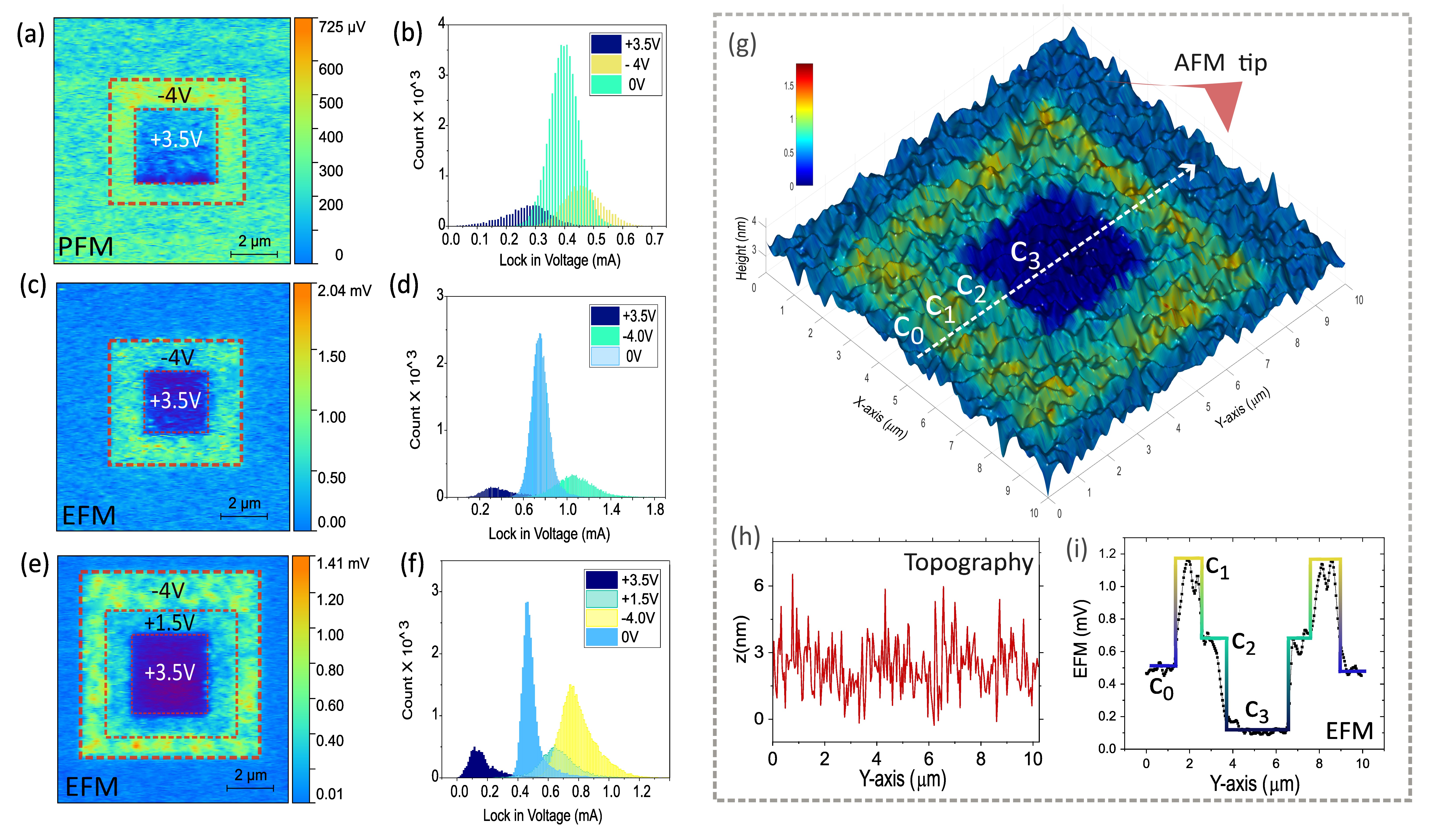}}
 \caption{ \textbf{Nanoscale evidence of multistate charge programmability.} (a) PFM phase mapping of the HZO film after box-in-box writing with $+3.5$~V and $-4$~V, revealing domain-level contrast. (b) The corresponding statistical distribution of piezoelectric response separates the written regions from the background, confirming two stable polarisation states despite low visual contrast. (c–d) EFM mapping of the same region shows stronger charge contrast and well-separated electrostatic charge regions in the statistical distribution. (e–f) EFM mapping of three states ($+3.5$, $+1.5$, and $-4$~V) demonstrates nanoscale multistate programmability. (g) Three-dimensional surface with EFM overlay. (h) The topography profile along the white dashed line in (g) shows a flat surface with RMS roughness $\sim$0.9~nm and no correlation with electrical contrast. (i) EFM line profile along the same line shows three distinct charge plateaus. 
}
\label{fig2}
\end{figure*}

To probe the microscopic origin of polarisation, we used Piezoresponse Force Microscopy (PFM) and Electrostatic Force Microscopy (EFM). PFM is a well-established tool to verify ferroelectricity at the nanoscale level, utilising the fact that polarised domains generate a piezoelectric response that can be mapped as a local mechanical signal \cite{kumar_multilevel_2025}. EFM, although not commonly used for ferroelectrics, provides complementary information by sensing electrostatic forces arising from bound charges and surface potentials, thereby directly linking domain orientation to stored charge \cite{fukuzawa_accurate_2023}. A box-in-box writing scheme was applied, with the outer region biased at \(-4\,\mathrm{V}\) and the inner region at \(+3.5\,\mathrm{V}\). As shown in Fig.~2(a), PFM resolves the two domains, where the \(-4\,\mathrm{V}\) write corresponds to the high-capacitance state (HCS) and the \(+3.5\,\mathrm{V}\) write to the low-capacitance state (LCS), as per electrical measurements. Achieving HCS requires a stronger field than LCS, in line with the observed electrical asymmetry. Even at \(-4\,\mathrm{V}\), the PFM contrast remains modest, likely due to screening from trapped charges and the relatively high PFM read drive, which can partially bias domains toward LCS. The histogram in Fig.~2(b) represents the statistical distribution of the piezoelectric response of the three regions, mitigating the weak contrast in Fig.~2(a) by separating the written regions from the unwritten region. It shows that the two written states occupy distinct PFM signal ranges and confirm stabilised remanent polarisation. To directly map capacitive switching through stored charge, EFM scans were performed on the same regions. As shown in Fig.~2(c-d), EFM provides stronger contrast because it senses electrostatic potentials arising from bound charges rather than piezoelectric strain. It maps two evident charge regions and verifies the non-volatility of states. To test intermediate states, three areas were written at \(-4\,\mathrm{V}\), \(+1.5\,\mathrm{V}\), and \(+3.5\,\mathrm{V}\) using the same box-inside-box protocol; the mid-level \(+1.5\,\mathrm{V}\) was chosen from electrical references to fall within the CMW. The EFM map and histogram in Figs.~2 (e) and (f) reveal three distinct charge levels, demonstrating nanoscale multistate storage, which is consistent with the capacitive results in Fig.~4(c). The EFM results were further verified to exclude cross-talk with surface topography. Figure~2(g) shows the 3D topography overlaid with the corresponding EFM measurement. A height line scan (Fig.~2h) appears smooth, while the EFM signal along the same path resolves three distinct charge plateaus (C1–C3), confirming that the contrast originates from trapped charges rather than surface morphology.

Together, the PFM and EFM measurements provide direct nanoscale evidence of binary and multistate polarisation configurations, with EFM demonstrating the first charge-resolved, non-volatile memcapacitive states in HZO films.

\subsection*{Device characterisation -- binary states}
Ferroelectric HZO capacitors exhibit characteristic polarisation–voltage (P-V) behaviour that shows two stable polarisation states. The thermodynamic origin of this bistable polarisation is described by Landau theory, which represents the Gibbs free energy ($F$) as a function of polarisation ($P$)\cite{clima_dielectric_2022,alam_transition-state-theory-based_2024}:

\begin{equation}
F(P) = \alpha P^2 + \beta P^4 + \gamma P^6 - EP
\end{equation}

Here, $\alpha$, $\beta$, and $\gamma$ are Landau coefficients and $E$ is the applied electric field. The resulting double-well energy profile contains two minima corresponding to remanent polarisation states ($P_{1}$ and $P_{2}$).  The switching occurs when an external field tilts the potential, driving domain nucleation and wall motion. Since the dielectric constant is inversely related to $\partial^2 F / \partial P^2$, each polarisation state modifies the permittivity, providing the physical basis for capacitive switching \cite{habibi_enhancing_2025}.
Figure~3(a) schematically illustrate the connection of this theoretical framework to experimental observables. Triangular voltage pulses are used to provide a continuous field sweep, allowing the P-V loop to be reconstructed from displacement currents. Below, it includes the theoretical prediction of the Landau model, along with the corresponding P-V and capacitive states. 

\begin{figure*}[t]
\centerline{\includegraphics[width=\textwidth]{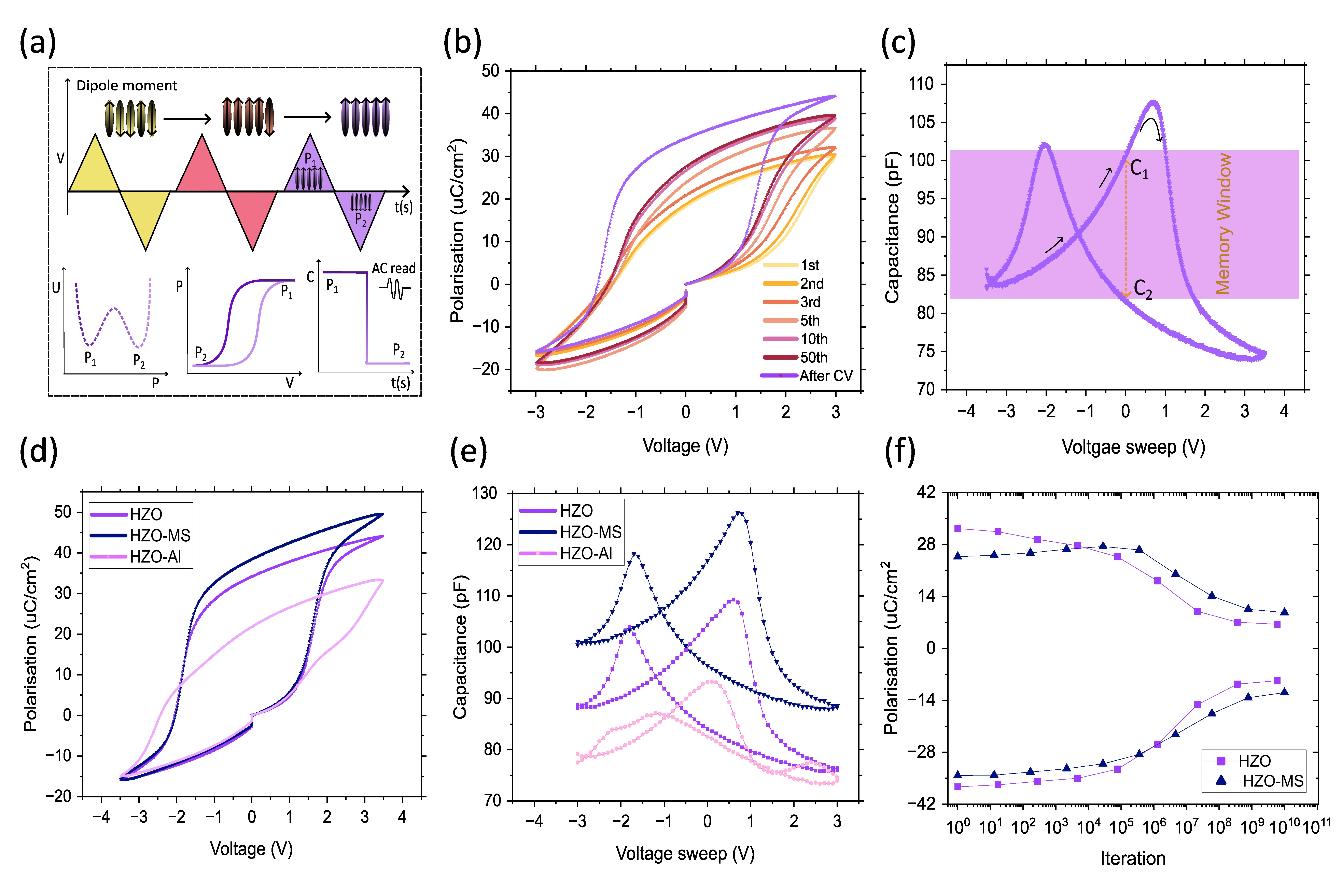}}
\caption{ \textbf{Binary switching and material stack optimisation}.(a) Schematic illustration of domain alignment required for polarisation using triangular pulses, associated double-well energy landscape, P-V and capacitance states. (b) P–V hysteresis of HZO shows progressive wake-up and stabilisation of remanent polarisation. (c) C–V loops from the same device display a butterfly shape with a 17~pF memory window and asymmetric coercive points, with an inset showing the two stable capacitance states. (d–f) Comparison of HZO, HZO-MS, and HZO-Al capacitors: (d) P–V loops reveal enhanced polarisation in HZO-MS and suppression in HZO-Al, (e) C–V shows an expanded window in HZO-MS, while (f) endurance measurements demonstrate improved cycling stability in HZO-MS. These results identify HZO-MS as the optimised stack for subsequent multistate studies.
}
\label{fig3}
\end{figure*}

Experimentally, two voltage protocols were employed to probe polarisation behaviour: large-signal triangular pulses for P-V measurements, and quasi-static small-signal sweeps for capacitance–voltage (C-V) measurements. P–V loops were recorded using 0.001\,s rise/fall times, as the response is strongly pulse-dependent (Fig.~2S). Figure~3(b) highlights the widely reported wake-up effect, where remanent polarisation gradually increases over $\sim$30–40 cycles, attributed to oxygen-vacancy redistribution, charge-trap depinning, and field-induced stabilisation of the metastable orthorhombic phase \cite{pesic_physical_2016, berruet_physical_2022}. Oxygen vacancies play multifaceted roles, contributing both to spontaneous polarisation and to leakage currents \cite{fang_electrical_2022, luo_highly_2020}. Interestingly, wake-up can be suppressed under C-V protocols, where the extended dwell time at each voltage step allows domains to align and stabilise, overcoming pinning effects. To confirm that the loops in Fig.~3(b) are of ferroelectric origin rather than leakage or interface artefacts, Positive-Up–Negative-Down (PUND) testing was performed (Fig.~3S(a)). Figure~3(c) shows the device exhibits an asymmetric butterfly-shaped C–V response, with a CMW of $\sim$20\, pF at zero DC bias. However, this window decreases after a certain number of cycles and switches to a $\sim$17\,pF window as shown in Fig.~3S(c). Capacitance maxima near the coercive fields reflect the differential polarisation response ($\mathrm{d}P/\mathrm{d}E$). Given the symmetric electrode stack, this asymmetry likely arises from interfacial phenomena such as oxygen vacancies and charge trapping, which generate internal bias fields and influence domain-wall pinning \cite{fang_electrical_2022, koduru_small-signal_2025, luo_non-volatile_2020}. The boundaries of the CMW define the high-capacitance state (HCS) and low-capacitance state (LCS): a negative sweep to $-3.0$\, V aligns domains into HCS, whereas a positive sweep to $+3.0$\,V programs LCS, as illustrated in the inset of Fig.~3(c).  As the readout is performed using a non-destructive small-signal AC probe (30\, mV) at zero DC bias, this binary capacitive switching with NDRO establishes a robust foundation for evaluating memory functionality.  

\subsection*{Material Stack Engineering}
The HZO capacitor demonstrates a reproducible capacitance memory window; however, the magnitude remains modest for advanced multistate applications. To address this, two modified stacks were fabricated and compared against the HZO device: (i) a multi-stack configuration (HZO-MS) with alternating 5:5 supercycles, and (ii) a bilayer stack incorporating a thin 1 nm layer of aluminium(HZO-Al), motivated by prior reports of polarisation enhancement \cite{lou_super-lamination_2024, li_record_2025, habibi_enhancing_2025}.
 The PV loops in Fig.~3(d) show that HZO-MS achieves a modest enhancement in remanent polarisation relative to the reference HZO. At the same time, HZO-Al exhibits a pronounced suppression, in contrast to earlier reports \cite{wu_effects_2024}. CV measurements Fig.~3(e) reinforce the conclusion that HZO-MS displays higher capacitance across the entire bias range and a larger non-volatile memory window ($\sim$24~pF) compared to standard HZO, whereas HZO-Al fails to exhibit a butterfly-shaped profile and loses its capacitive memory functionality reported before \cite{si_ferroelectric_2019}. Finally, endurance testing in Fig.~3(f) confirms that HZO-MS retains stable operation over $10^6$ cycles, moderately outperforming the reference HZO and far exceeding the degraded HZO-Al response (see Fig.~S4). 
 Among the three stack designs, the HZO-MS device consistently outperforms the standard HZO and HZO-Al stack, making it the optimal platform for multibit characterisation.
 
\begin{figure*}[t]
\centerline{\includegraphics[width=\textwidth]{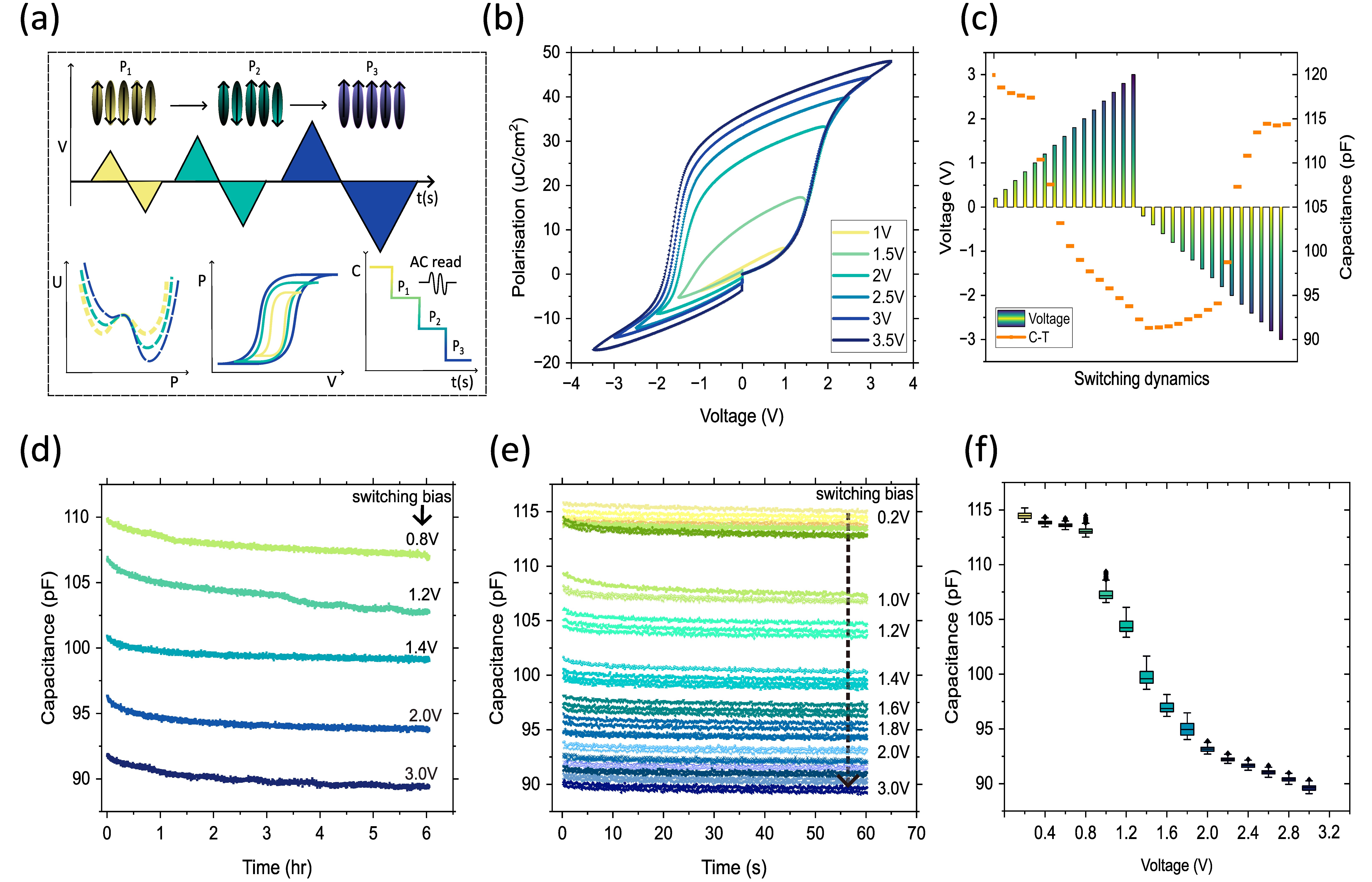}}
\caption{ 
\textbf{Multistate tuneable memcapacitor.} 
(a)  Triangular voltage pulses of increasing amplitude (colour-coded) progressively orient the dipoles, tilting the Landau double-well energy landscape as predicted by theory. The corresponding $P$–$V$ and $C$–$t$ schematics highlight how intermediate polarisation states map onto multiple stable capacitance states.
(b–f) Experimental results on HZO-MS devices. 
(b) polarisation–voltage loops under increasing amplitude, revealing quasi-stable intermediate configurations. 
(c) Bidirectional capacitive switching dynamics showing asymmetric potentiation and depression. 
(d) Retention of five capacitance states over 6 h. 
(e) Evolution of capacitance under repeated programming at fixed bias. 
(f) Statistical distribution of capacitance values versus programming voltage, with variability highest in mid-window states.
}
\label{fig4}
\end{figure*}

\subsection*{Tuneable Multistate Capacitance} 
Multistate behaviour in ferroelectrics arises naturally from the Landau energy landscape. Varying the amplitude of the applied voltage tilts the double-well potential, enabling the system to stabilise in intermediate quasi-stable configurations. Instead of switching fully between the two extreme polarisation states, multiple intermediate states can thus be accessed. This mechanism is illustrated schematically in Fig.~4(a), where increasing voltage amplitude progressively reorients dipoles, corresponding to distinct energy minima and polarisation levels that map onto multiple capacitive states and have been reported in earlier studies \cite{zhang_flexible_2025,xie_ferroelectric_2023,kumar_multilevel_2025,kim_switching_2022}. To translate this phenomenon into our HZO-MS devices, we applied progressive voltage pulses, which produced a corresponding increase in remanent polarisation, as shown in Fig.~4(b). To map this into capacitance states, sequential positive voltage sweeps of increasing amplitude were applied, causing the device to switch from $\sim$118~pF to 92~pF and form a ladder of 8-9 capacitive states. Each of these states is non-volatile and read out using only a 30 mV AC probe, enabling NDRO.
Reversing the sweep polarity restores the system to the HCS through analogous steps, confirming fully bidirectional reversibility [Fig.~4(c)]. The switching, however, is asymmetric, mirroring the C–V hysteresis observed in Fig.~3(c). Positive sweeps initiate near 1~V and access multiple intermediate states, whereas negative sweeps require stronger fields (onset $\sim$–1.6~V) and evolve more gradually toward saturation. This gradual potentiation under negative bias, accompanied by sharper depression under positive bias, offers a close analogue to synaptic weight modulation for neuromorphic computing \cite{tian_ultralowpower_2023}. In addition, retention tests over 6 h confirm that all intermediate states remain well separated, with early relaxation before long-term stabilisation [Fig.~4(d)].
Beyond the functional asymmetry, repeated cycling reveals a modest narrowing of the CMW from $\sim$26 pF to $\sim$24 pF in Fig.~4(e), a behaviour typically associated with domain equilibration. 

Multistate switching was further evaluated under repeated programming at fixed voltage amplitudes. In this protocol, the same bias was applied successively until the capacitive response saturated, revealing both how much of the memory window a given voltage can access and whether this behaviour is consistent. As shown in Fig.~4(e), successive sweeps at a defined voltage gradually reduce the capacitance before reaching a stable state, indicating continued domain reorientation with each write. Thus, a single programming voltage does not correspond to a unique capacitive state, but instead unlocks a spectrum of quasi-stable states, with the final capacitance being governed by both the bias magnitude and the write history. Figure~4(f) quantifies this behaviour by plotting the distribution width of remanent capacitance versus voltage. The spread is most extensive in the mid-window region, where domain activity is most pronounced, and narrows toward the extremes, where polarisation approaches saturation. This variability sets the effective resolution of multistate operation. It underscores that ferroelectric capacitors can be reproducibly programmed into history-dependent configurations, providing a robust physical basis for tunable circuit elements.

\begin{figure*}
\centerline{\includegraphics[width=\textwidth]{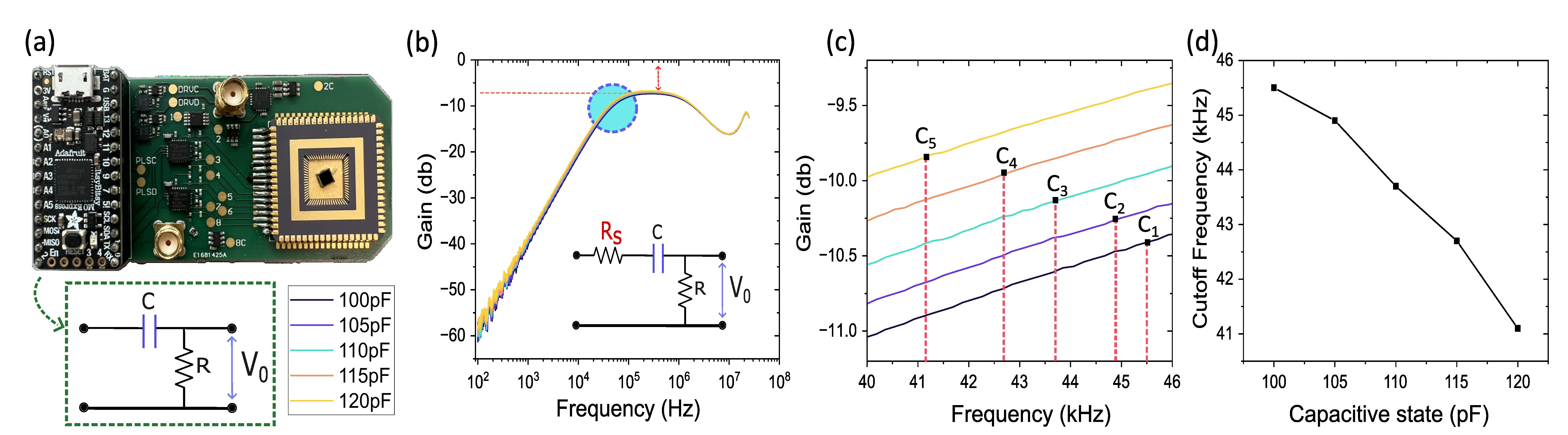}}
\caption{\textbf{Tuneable high-pass filtering with ferroelectric memcapacitor. } 
(a) Packaged HZO-MS capacitor integrated on a PCB, together with the equivalent RC circuit and the five programmable capacitance states ($C_1$–$C_5$). A small systematic offset of $\sim$5~pF is observed between wafer-scale and packaged devices, attributed to parasitic contributions. (b) Measured filter response compared to the ideal RC model. The reduced cutoff frequency and depressed pass-band gain indicate the presence of an additional series resistance $R_S$. (c) Magnified view near the cutoff region showing five clearly separated frequency responses, each corresponding to a distinct capacitance state. (d) Extracted cutoff frequency as a function of programmed capacitance. 
}
\label{fig5}
\end{figure*}
\subsection*{Tuneable High-pass filter}
To demonstrate circuit-level utility of the capacitive switching, the packaged ferroelectric memcapacitor chip was mounted on a PCB and used as the tuneable element in a simple RC high-pass filter with a 15~k$\Omega$ resistor as shown in Fig.~5a . Five non-volatile capacitance states spanning $\sim$100--120~pF were programmed and verified on a source-measure unit. The packaged device shows a $\sim$5~pF upward shift relative to wafer measurements, consistent with parasitic contributions from wiring and the substrate. Based on $f_c = 1/(2\pi RC)$, the targeted cutoff band for this range is 106-87~kHz. However, the experimental magnitude response in Fig.~5b exhibits a lower cutoff of $\sim$40~kHz and a pass band near $-7$~dB rather than the ideal 0~dB. The depressed pass band and lowered cutoff suggest the presence of an additional series resistance, which is difficult to detect in conventional source-measure characterisations and therefore remained non-essential until revealed at the circuit level. This indicates that the circuit should be modeled with an extra $R_s$. Some contribution from dielectric loss may also be present, dissipating signal power across the bandwidth. 

While the overall response in Fig.~5b shows the band edges, the individual states are difficult to observe due to their close spacing. A zoomed-in view near the cutoff (Fig.~5c) clearly resolves five well-separated states, demonstrating that each programmed capacitance translates directly into a distinct and reproducible cutoff from 45.5 to 41.1~kHz, with a relative tuning of $\sim$9.7\% achieved without any static bias. Ideally, the tuning range should span from 50 to 41.1~kHz, following the expected 106/87~$\approx$~1.21 ratio. The observed narrowing of this window can be attributed to variations in $R_s$ with programming voltage. As the capacitance increases, the pass-band gain approaches 0~dB, suggesting a decrease in $R_s$ (values summarised in Table~1). In contrast, at lower capacitances, that is, the higher cutoff side, $R_s$ is larger, effectively pinching the accessible window from the expected $\sim$50~kHz down to $\sim$45~kHz. Figure~5(d) further supports the presence of non-idealities, as the variation of measured cutoff frequency with capacitance deviates from the expected $f_c \propto 1/C$. Presence of non-idealities has previously been reported in other memcapacitor stacks, suggesting a path unexplored for ferroelectric memcapacitor \cite{yadav_impedance_2025}. Here, it highlights $R_s$ as a hidden yet quantifiable circuit-level factor in ferroelectric capacitors, offering a new pathway to probe device loss mechanisms. Crucially, all states are non-volatile and reversible under repeated programming, establishing that multistate ferroelectric capacitors can deliver stable, reprogrammable filter responses at the PCB level.

\section*{Conclusion}
We demonstrated that ferroelectric HZO memcapacitors can move beyond binary polarisation switching to achieve stable, multilevel capacitance states directly exploitable at the circuit level. The optimised stack exhibited non-volatile, voltage-controlled switching across 8–9 reprogrammable states, with an improved capacitive memory window of $\sim$23~pF. These states remained stable under zero bias, demonstrating retention beyond 6~h and endurance exceeding $10^6$ cycles. Nanoscale validation EFM confirmed the underlying charge-based switching dynamics. At the circuit level, integration into a high-pass filter yielded reprogrammable, bias-free frequency responses spanning a $\sim$5~kHz window. This achievement marks a significant advance in tunable capacitors, positioning HZO memcapacitors as compact, low-power, and CMOS-compatible building blocks for adaptive RF filters, reconfigurable analogue hardware, and neuromorphic electronics.

\section*{Methods}

\subsection*{Fabrication}
Metal-ferroelectric-metal standalone capacitors were fabricated on a 200~nm silicon dioxide (SiO$_\text{2}$) layer, grown by dry oxidation at 1050$^\circ$C on a silicon substrate. For the top and bottom electrodes, a 50~nm TiN layer is deposited using an Angstrom sputtering system at room temperature, working pressure of 2~mTorr, 600~W power, and 20~sccm nitrogen gas flow. A 10~nm platinum capping is sputter-coated to protect the top and bottom TiN electrodes from RTA. The ferroelectric layer is deposited using thermally enhanced atomic layer deposition (ALD) with tetrakis(dimethylamido)hafnium (TDMAHf), Tetrakis(dimethylamido)Zirconium (TDMAZ), and Trimethylaluminium (TMA) as the hafnium, zirconium, and aluminium precursors, respectively, and water as the co-reactant. HZO consists of consecutive cycles of HfO$_\text{x}$:ZrO$_\text{x}$, HZO-MS uses alternating sets of 5 cycles of HfO$_\text{x}$ and ZrO$_\text{x}$, and HZO-Al with an additional 1 nm Al$_\text{2}$O$_\text{3}$ layer deposited in the middle of the HZO layer. Devices with lateral dimensions of 20$\times$20, 40$\times$40, and 60$\times$60~$\mu$m$^2$ were defined photolithographically using a KS mask aligner and patterned by a combination of lift-off and reactive ion etching with JLS. Post-fabrication, RTA is performed at 500$^\circ$C for 30~s in an ambient nitrogen environment to induce crystallisation and promote the formation of the orthorhombic phase.

\subsection*{Material Characterisation}
A Panalytical Empyrean XRD at grazing incidence with Co K-$\alpha$ radiation ($\lambda$=1.7886) is used to investigate the crystallinity of the film for mono, tetra and orthorombic phases. For XPS, a Thermo Scientific NEXSA G2, equipped with a monochromated Al k-$\alpha$  X-ray source (1486.7~eV) is used to check the film stoichiometry. A gentle surface cleaning was performed using an argon cluster source (4 keV, 1000 atoms) for 30 seconds to minimise contribution from carbon. PFM and EFM measurements were performed using a Park Systems NX20, equipped with a conductive PPP-CONTSCPt probe. PFM readout was conducted using an AC driving voltage of 10\,V at 100\,kHz to detect the out-of-plane piezoresponse. In comparison, EFM read were carried out with 30\,mV AC signal at 17\,kHz to probe surface charge distributions under ambient conditions. SEM imaging of the devices was performed using a TESCAN VEGA3.

\subsection*{Electrical Characterisation}
A Keithley 4200 Semiconductor Characterisation System parameter analyzer is used to conduct all ferroelectric measurements. All measurements are from the 60$\times$60~$\mu$m$^2$ device, to achieve maximum polarisation as explained in Fig.~3S(b). Retention measurements were performed using a readout delay of 1 minute, followed by a read every minute for 6 hours. Endurance pulse parameter are as follows,  pulse width: $10^{-5}$s, delay time: $10^{-5}$~s, rise/fall time: $10^{-6}$~s, and 100 points per waveform. The endurance test is configured with an iteration size of 100, fatigue count of 10, and a maximum loop count of $10^{9}$, resulting in a total of $10^{10}$ cycles. All electrical measurements are carried out at room temperature under ambient conditions. For circuit-level characterisation, frequency response measurements of the RC high-pass filter were carried out using an R\&S RTB2000 oscilloscope. Frequency sweeps were performed from 100~Hz to 25~MHz with 500 points per decade and a 30~mV peak-to-peak input amplitude to capture both cutoff behaviour and pass-band response.

\section*{Acknowledgments}
We thank the Cleanroom Facility at the University of Glasgow for support with RTA. At the University of Edinburgh, Gus Calder (School of Geosciences) provided XRD measurements, and Graham Wood (Scottish Microelectronics Centre) supported device dicing and wire bonding. We also acknowledge the wider support of the Scottish Microelectronics Centre facilities.

\section*{Author Contributions}
D.Y. conceived and carried out the device fabrication, electrical characterisation, analysis, filter testing, and manuscript preparation. S.S. provided continuous support for device data analysis, including setting up the multiplexer for filter testing. P.F. designed the PCB and assisted with filter measurements. M.A. and H.L. supported AFM and XPS measurements and their analysis. A.T. assisted with SEM imaging and resolved contact pad issues for wire bonding. T.P. contributed to the direction, discussions, and critical feedback.

\section*{Competing Interests}
The authors declare no conflict of interest.

\bibliographystyle{unsrt}
\bibliography{references-2}

\end{document}


\maketitle
\section*{Supplementary Figures}
\renewcommand{\thefigure}{\arabic{figure}S}

\begin{figure}[H]
    \centering
    \includegraphics[width=\textwidth]{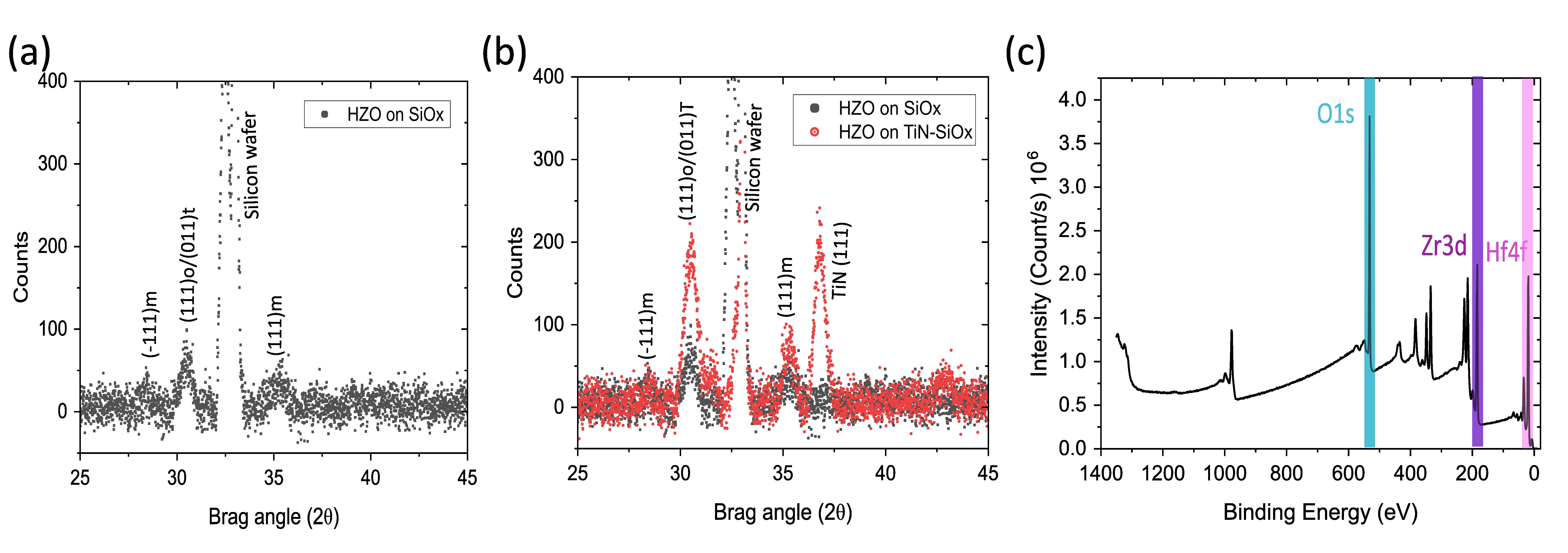}
    \caption{\textbf{Structural and chemical analysis of HZO films.} 
(a) XRD of HZO on SiO$_x$, showing the coexistence of monoclinic and tetragonal/orthorhombic phases \cite{vogel_structural_2022}. A sharp peak from the Si wafer substrate is visible near 32°, while the feature near 37° arises from TiN, as confirmed in (b). (b) Comparison of HZO on SiO$_x$ and on TiN/SiO$_x$. The additional reflection at $\sim$37° corresponds to the TiN electrode \cite{article}, confirming its contribution, while the orthorhombic feature is enhanced by the presence of TiN. (c) Survey XPS spectrum of HZO film, highlighting the O~1s, Zr~3d, and Hf~4f core levels. The comparable intensities of the Zr and Hf peaks confirm an approximate 1:1 Hf:Zr atomic ratio, consistent with optimised ferroelectric HZO compositions.
}
\label{Fig1S}
\end{figure}

\begin{figure}[H]
    \centering
    \includegraphics[width=\textwidth]{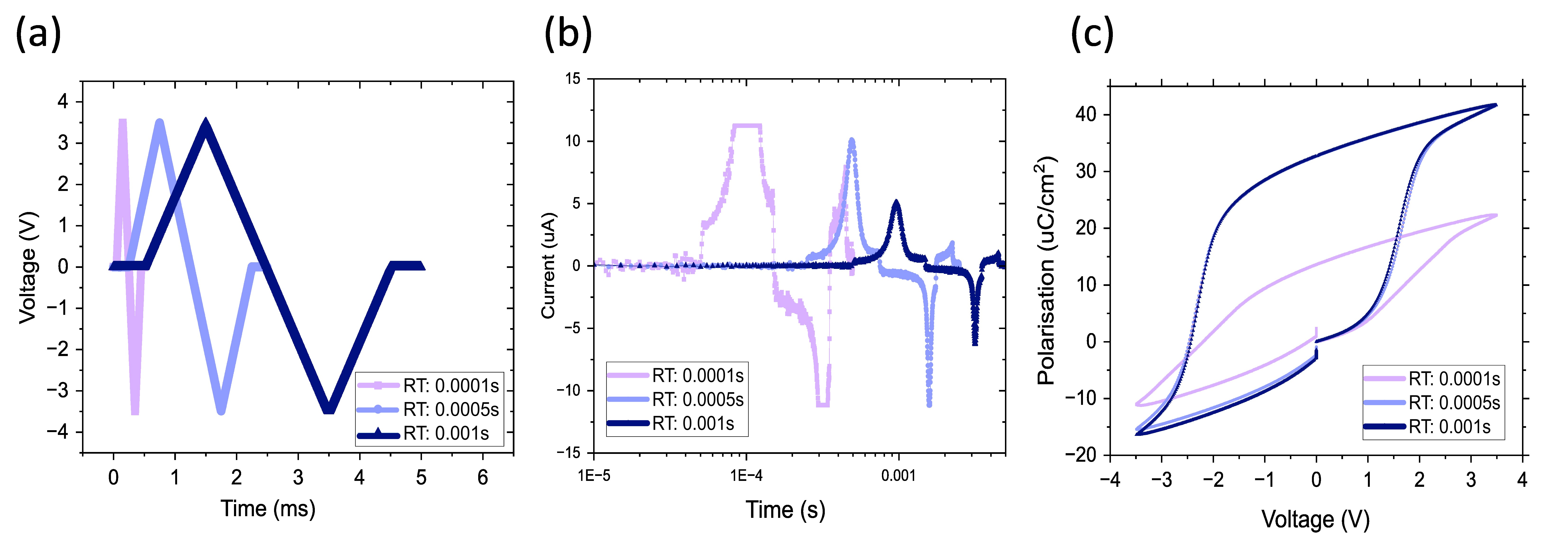}
    \caption{\textbf{Effect of triangular pulse rise and fall time on ferroelectric switching}
(a) Input triangular waveforms with rise/fall times of 0.0001, 0.0005, and 0.001 s. (b) Corresponding displacement current transients show strong dependence on ramp rate: very short rise/fall times (0.0001 s) produce distorted peaks, whereas slower ramps reduce peak amplitude.
(c) The polarisation–voltage (PV) loops reveal that despite large differences in displacement current, the loops for 0.0005 and 0.001 s are nearly identical, confirming that once domain switching is saturated, the integrated polarisation (total charge) remains constant. While the current peak decreases with slower ramps, the longer duration compensates, conserving charge flow and yielding consistent polarisation values. These results indicate that rise/fall times above 0.0005 s are sufficient to achieve maximum polarisation.}
\label{Fig2S}
\end{figure}

\begin{figure}[H]
    \centering
    \includegraphics[width=\textwidth]{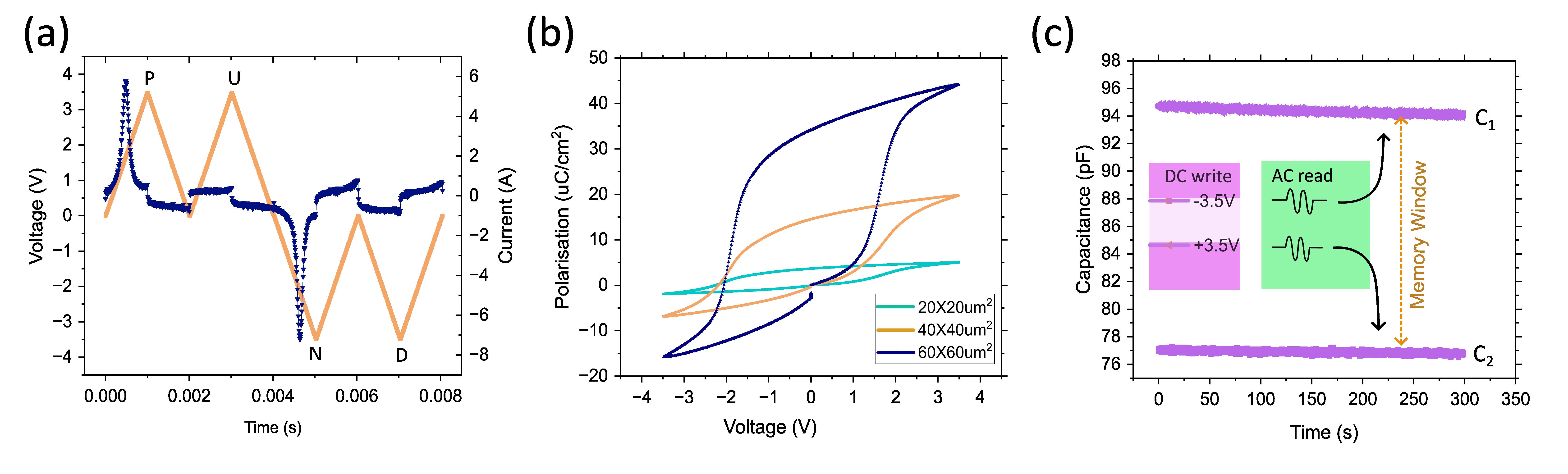}
    \caption{ (a) PUND measurement showing the applied triangular waveform and the corresponding displacement current. 
PUND is essential to isolate the switchable ferroelectric polarisation from non-switchable contributions such as leakage and dielectric charging, thereby confirming the intrinsic ferroelectric response. 
The displacement current peaks correspond to the P and N pulses and disappear upon repetition, indicating that the signal originates from ferroelectric domain switching rather than dielectric charging. 
In contrast, the dielectric current is visible in the consecutive U and D pulses \cite{magagnin_novel_2025}.  
(b) polarisation–voltage (PV) loops of capacitors with different device areas (20$\times$20, 40$\times$40, and 60$\times$60~$\mu$m$^2$), confirming proper area scaling of polarisation.  
(c) In reference to the memory window in Fig.~3(c), demonstration of non-volatile capacitive states. Two stable capacitance levels ($C_1$ and $C_2$) are written using $\pm$3~V DC programming pulses and read using a 30~mV AC signal at zero bias, confirming robust NDRO (non-destructive readout).  The memory window is comparatively less than in the first few run, as dicussed in Fig. 4(e). 
}
\label{Fig3S}
\end{figure}

\section*{Supplementary Table}

\begin{table}[H]
    \centering
    \caption{Comparison of expected and measured cut-off and passband values}
    \begin{tabular}{@{}ccccc@{}}
        \toprule
        Capacitance (pF) & Measured $f_c$ (kHz) & Expected $f_c$ (kHz) & Pass Band (dB) \\
        \midrule
        120 & 41.1 & 88.5 & -6.85 & \\
        115 & 42.7 & 92.3 & -6.96 & \\
        110 & 43.7 & 96.2 & -7.1 & \\
        105 & 44.9 & 101.3 & -7.3 & \\
        100 & 45.5 & 106.1 &  -7.4 & \\
        \bottomrule
    \end{tabular}
    \label{tab:cutoff-passband}
\end{table}

\bibliographystyle{unsrt}
\bibliography{references}